\documentclass[conference]{IEEEtran}
\IEEEoverridecommandlockouts
\usepackage{cite}
\usepackage{multirow}
\usepackage{amsmath,amssymb,amsfonts}
\usepackage{algorithmic}
\usepackage{graphicx}
\usepackage{textcomp}
\usepackage{diagbox}
\usepackage{url}
\usepackage{xcolor}
\def\BibTeX{{\rm B\kern-.05em{\sc i\kern-.025em b}\kern-.08em
    T\kern-.1667em\lower.7ex\hbox{E}\kern-.125emX}}
\begin{document}
\graphicspath{{figures/}}

\title{Early Detection of Network Attacks Using Deep Learning
	\thanks{Pre-print submitted to ITEQS 2022}
}

\author{\IEEEauthorblockN{Tanwir Ahmad\IEEEauthorrefmark{1}, Dragos Truscan\IEEEauthorrefmark{1}, Jüri Vain\IEEEauthorrefmark{1}\IEEEauthorrefmark{2}, and Ivan Porres\IEEEauthorrefmark{1}}
	\IEEEauthorblockA{\IEEEauthorrefmark{1}\textit{Faculty of Science and Engineering, \AA bo Akademi University, Finland} \\
	\IEEEauthorrefmark{2}\textit{High-assurance Software Laboratory, Tallinn University of Technology, Estonia} \\
		\{tanwir.ahmad, dragos.truscan, ivan.porres\}@abo.fi, juri.vain@ttu.ee}
}

\maketitle

\begin{abstract}

The Internet has become a prime subject to security attacks and intrusions by attackers. These attacks can lead to system malfunction, network breakdown, data corruption or theft.
A network intrusion detection system (IDS) is a tool used for identifying unauthorized and malicious behavior by observing the network traffic.
State-of-the-art intrusion detection systems are designed to detect an attack by inspecting the complete information about the attack. This means that an IDS would only be able to detect an attack after it has been executed on the system under attack and might have caused damage to the system.
In this paper, we propose an end-to-end early intrusion detection system to prevent network attacks before they could cause any more damage to the system under attack while preventing unforeseen downtime and interruption. 
We employ a deep neural network-based classifier for attack identification. The network is trained in a supervised manner to extract relevant features from raw network traffic data instead of relying on a manual feature selection process used in most related approaches.
Further, we introduce a new metric, called earliness, to evaluate how early our proposed approach detects attacks.
We have empirically evaluated our approach on the CICIDS2017 dataset. The results show that our approach performed well and attained an overall 0.803 balanced accuracy.

\end{abstract}


\section{Introduction}

Modern society is significantly dependent on a wide range of inter-connected software systems for finance, energy distribution, communication, and transportation. The era of controlled communication in closed networks for restricted purposes is over. 
Due to the adoption of Internet technologies, almost all financial, government, and social sectors started to rely heavily on networked information systems to process and store confidential information. 
As a result, these systems have become primary subjects to security attacks and intrusions by attackers. These attacks can lead to system malfunction, network breakdown, data corruption or theft.
Therefore, it is essential to ensure network security by monitoring and detecting network attacks in real-time as early as possible.

A network \textit{intrusion detection system} (IDS) is a tool used for identifying unauthorized and malicious behavior by observing the network traffic and helping network administrators take appropriate preventive measures to secure the network infrastructure and the associated nodes~\cite{Mukherjee94}. The majority of the intrusion detection systems can be divided into two groups: \textit{anomaly-based} and \textit{signature-based} detection systems~\cite{Ahmad2021}. 
In the former group, a detection system learns the profile of normal network traffic and would classify the given network traffic data as intrusive or anomalous if it deviates from the normal traffic profile by more than a pre-defined anomaly threshold. 
This allows these systems to detect undiscovered and novel attacks.
However, the value of the anomaly threshold has a significant impact on the accuracy of the systems. Finding the optimal value of anomaly threshold is a complicated task and, in many cases, requires manual tuning.

A \textit{signature-based} intrusion detection system identifies intrusive network traffic by comparing the given network traffic data against the signatures (e.g., sequence of string and regular expressions) of known attacks. 
This category of network IDS  is most commonly used in daily practice \cite{Vigna2004}.
Since most of these systems rely on the knowledge bases (i.e., pre-defined sets of attack models and patterns) extracted from known attacks and system vulnerabilities, they are also known as \textit{knowledge-based} or \textit{misuse} IDS~\cite{Liao2013}.
In most cases, the domain experts construct the knowledge bases manually, which can be a tedious and error-prone task~\cite{Li2019}.
Unlike anomaly-based IDSs, this group of IDSs can only detect those attacks that are defined in the knowledge bases. 
However, these methods demonstrate high degree of accuracy and a low false alarm rate compared to the anomaly-based IDSs~\cite{Ahmad2021, GarciaTeodoro2009}. 
One of the main challenges in developing these systems is extracting or defining a signature of a known attack that can represent different variations of the attack. Furthermore, managing a large signature knowledge-base and matching signatures against the traffic are time and resource-intensive tasks~\cite{Ahmad2021}.

The spread of high-speed networks and fast-propagating threats pose additional challenges to current IDSs, which detect an attack by inspecting the entire network traffic data related to the attack. This means that an IDS would only be able to detect an attack after it has been executed on the system under attack and might have already caused damage to the system.
Therefore, early attack (or intrusion) detection is desirable in the cyber-security domain to prevent network attacks before they could cause any more damage to the system.
Based on the early classification results of the traffic, the network administrator can decide whether to stop the traffic and raise an alarm message or deploy counter measures.

In order to address the above challenges, in this work, we propose an end-to-end early signature-based IDS to detect network attacks as early as possible with a high degree of accuracy. However, instead of manually extracting the signatures of attacks, our method employs a \textit{Deep Neural Network} (DNN) in order to extract relevant features from raw network traffic data that can be used for early classification of ongoing attacks. Deep learning is a type of machine learning where we utilize DNNs~\cite{Bengio12a} or multi-layer neural networks to approximate complex functions by learning different levels of representations of the given training data.
In summary, the contributions of this paper are as follows:
\begin{enumerate}
	\item We present an early network intrusion detection method. Intuitively, our approach would be able to take a more informed decision about the class label (e.g., normal or malicious) of a given network traffic data if more data is available, but it will delay the decision. Thus, in this work, we focus on optimizing the accuracy of attack detection with minimum delay (or maximum earliness).
	
	\item Our approach extracts the relevant features from raw network traffic data in an end-to-end manner instead of relying on the manual feature engineering process. Therefore, our approach is domain-independent and does not require domain specific data pre-processing steps.
	
	\item We introduce a new metric, called \textit{earliness}, to evaluate how early our proposed approach can detect attacks.
	
	\item We provide the tool support for our approach using the Keras~\cite{chollet2015keras} library.
	
	\item We empirically evaluate our approach using the CICIDS2017~\cite{Sharafaldin2018a} dataset.
	
\end{enumerate}

The rest of the paper is structured as follows: 
Section \ref{sec:approach} describes our approach.
In Section \ref{sec:evaluation}, we empirically evaluate our approach. Section \ref{sec:related-work} presents an overview of the related work.
Section \ref{sec:threats-to-validity} specifies threats to the validity of this work. Finally, Section \ref{sec:conclusion} draws conclusions and discusses future work.

\section{Approach}\label{sec:approach}
\begin{figure*}[htbp]
	\centering 
	\includegraphics[width=\textwidth, trim={1.5cm 0 1.5cm 0},clip]{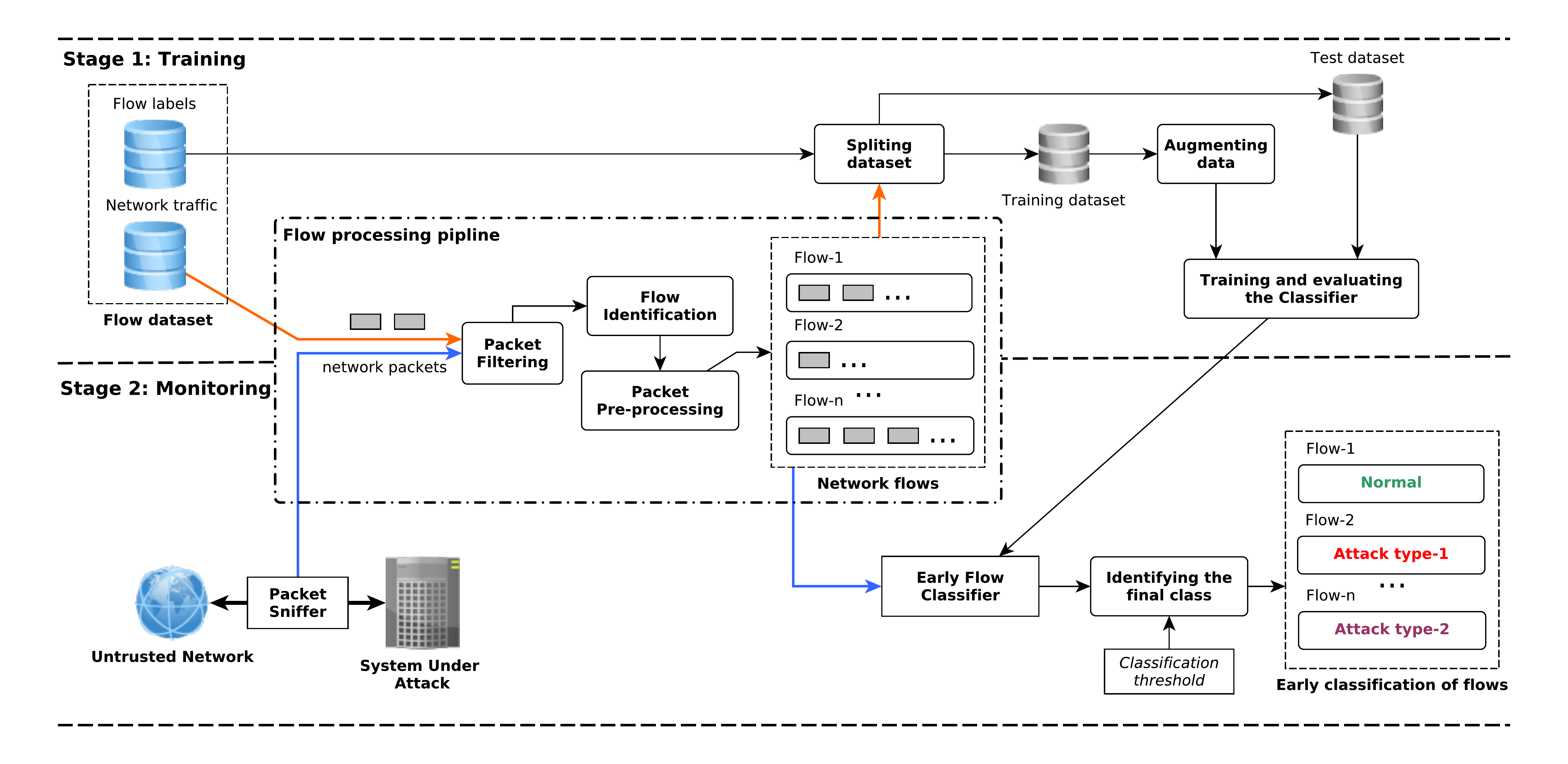}
	\caption{Stages of our approach}
	\label{fig:approach}
\end{figure*}

This section proposes an early intrusion detection system for identifying network attacks.
The main objective of our approach is to monitor the network traffic in real-time, automatically extract the features from raw network traffic data, avoid time-consuming and laborious task of extracting the features using traditional methods, and accurately detect the network attacks as early as possible.
Our approach can be divided into two stages, as shown in Figure~\ref{fig:approach}. At the first stage, we train and evaluate our early flow classifier using a dataset containing labeled network flows and network packets corresponding to those flows. At the second stage, we employ the trained early flow classifier to predict whether a given network flow is likely to be malicious or normal. 
A \textit{network flow} is a bidirectional sequence of packets exchanged between two endpoints (e.g., a web server and a client) during a certain time interval with some common flow properties \cite{Claise2013SpecificationOT} such as source and destination IP addresses, source and destination port numbers, and the protocol type. In our work, we define a network flow as a sequence of \textit{T} ordered packets, where \textit{T} represents the length of a complete flow. A flow is denoted as:
\begin{align} \label{eq:flow}
	F_T=\{P_{1}, P_2,...,P_T\} \; \forall \; P_i \in \mathbb{R}^d \wedge 1 \leq i \leq T
\end{align}
where \textit{d} is the dimension (or length) of a packet.

Both stages of our approach use the \textit{flow processing pipeline} that consists of three modules: packet filtering, flow identification, and packet pre-processing module.
The packet filtering module captures the network packets and forwards them to the subsequent modules if they satisfy the given criteria. For example, if we are protecting the web server running at port 80, we can configure the module to forward only those packets whose destination or source port is 80.
The next two modules transform the packets and group them into network flows. Whenever a network flow is updated with a new packet, we use the early flow classifier to update the prediction corresponding to the flow.

\subsection{Flow Processing Pipeline}\label{sec:flow-processing-pipeline}
In this section, we will discuss the modules of the pipeline.

\subsubsection{Packet Filtering}\label{sec:packet_filter}
We monitor the raw network traffic between the system under attack and the untrusted network. We select only those network packets which are related to the type of attacks we would like to detect. For example, if we are interested in detecting only web attacks~\cite{ashwini2013}, we will capture only HTTP packets.

\subsubsection{Flow Identification}\label{sec:packet-grouping}
Upon receiving a new packet, we inspect the packet properties such as source and destination IP addresses to identify a suitable active flow for it.
An \textit{active flow} represents an ongoing communication session between a pair of network endpoints.
On the other hand, if we cannot find an active flow that matches the characteristics of a packet, we create a new flow.
A network flow is considered to be \textit{terminated} or \textit{inactive} upon connection teardown (e.g., by FIN packet) or when the flow has not received a new packet within a certain flow timeout (e.g., 120 seconds). The flow timeout value can be adjusted according to the protocol type of the network traffic we are capturing for detecting attacks.

\subsubsection{Packet Pre-processing}\label{sec:packet-pre-processing}
Once we have identified the appropriate flow for the new packets, each packet goes through the following pre-processing steps to truncate unwanted information and transform it to a uniform-size vector of bytes: truncation and transformation. The main purpose of the steps is to ensure that the classifier should rely on relevant features for flow classification.
In this paper, we discuss the steps in relation to HTTP and TCP protocol; however, these steps can be applied to other types of network packets with minor modifications.

\paragraph{Truncation} The raw captured packets contain the Ethernet header. The header has information concerning the physical link, such as the Media Access Control (MAC) address used for transferring the frames between different nodes in the network. However, this information is valueless for attack identification because it can be spoofed easily. Thus, this header is removed from the packet.

Similarly, the Internet Protocol (IP) header in the packets includes information such as the total length of the packet, protocol version, source and destination IP addresses. This information is necessary for routing packets in the network. However, we consider this information irrelevant and counter-productive for our classifier since there is a chance that the classifier will start relying on the IP information (e.g., IP addresses) for detecting attack flows. Therefore, we remove it from the packets.
This allows the classifier to function steadily even if the addresses of the nodes in the network have changed and generalize the knowledge learned from one network environment to another.

\paragraph{Transformation} 
A fixed-size input is required when using a neural network for classification.
To make the length of the header of the Transport layer and the payload of the packets uniform, we crop or pad them with zeros to a fixed length.  
We would like to point out that, even though we restrict the length of the packets (i.e., \textit{d} in Equation \ref{eq:flow}) in a flow; we do not restrict the length of a flow (i.e., number of packets \textit{T}) unlike other proposed approaches (e.g., \cite{Zhang2019}) though it is implicitly bounded by time out.

\subsection{Early Flow Classification}\label{sec:early-flow-classification}
In order to make prediction actionable, one essential aspect is to predict an attack with adequate lead time so that the correct countermeasures against the attack can be enforced as quickly as possible before it could cause any more damage to the system under attack. 
Therefore, in order to minimize the prediction time, we employ a relatively lightweight DNN as an early flow classifier because as the size (i.e., number of trainable parameters) of a model grows, the prediction time increases as well since more computations are required to make a prediction. However, the major problem is that a reduction in the model size typically leads to limited expressive power and to low accuracy \cite{cloudArchitecture,Lu2017}. In order to mitigate this problem, instead of training one large or complex model to classify all types of attacks, we train a collection of simple models where each model is trained to classify only a subset of attack classes (e.g., 3 to 4 classes) of interest. There are several ensemble strategies proposed in the literature such as majority voting and ranking to employ multiple models and combine their predictions \cite{Yang2004}. However, it is out of the scope for this paper to discuss ensemble strategies. We leave this as our future work.

In this work, we use one-dimensional Convolutional Neural Networks (ID-CNNs)~\cite{Lecun1998} (i.e., a type of DNN) to extract a good internal representation of network flows and provide it as an input to a fully-connected or dense layer. We use a \textit{softmax} layer \cite{Goodfellow2016} as the final layer of our network to calculate a probability distribution for target classes.
The CNNs are used to extract relevant features from grid-shaped input data such as images and sequences. They are well capable of modeling the spatial and temporal dependencies in the data by learning relevant convolution filters (i.e., a set of grid-shaped weights or trainable parameters). A convolution layer is composed of several convolution filters where each filter is used to extract a certain feature from the input data. Thus, the output of a convolution layer is called a \textit{feature map}. 

The input data for ID-CNNs has two dimensions. The first dimension specifies the sequence of events (i.e., packets in a network flow); whereas, the second dimension correlates to the individual features of an event (i.e., bytes of a packet). We have used \textit{Rectified Linear Unit} (ReLU)~\cite{Nair2010} as a nonlinear activation function for every neuron in the convolutional layer. Typically, each convolutional layer is followed by a pooling layer~\cite{Goodfellow2016} to achieve translation invariance of the output returned by the convolutional layer. This layer reduces the temporal size of the output by replacing each fixed-size partition of it with a summary statistic (e.g., maximum or average) of the adjacent elements. 
The CNNs have a smaller number of trainable parameters than other types of artificial neural networks such as fully-connected networks~\cite{Lecun1998}. Therefore, they are less likely to overfit the training data than the fully-connected networks results in a better generalization.

After convolution and pooling operations, a given variable-length network flow is represented by a variable-length series of feature maps. We use a \textit{global pooling layer}~\cite{lin2014network} to transform the series into a fixed-length vector, which is then provided as an input to fully connected layers to get the feature vector. Lastly, we apply a \textit{softmax} layer to the feature vector in order to obtain a probability distribution for each class. Based on the probability distribution and the classification threshold, we make the final predictions.

\subsection{Training}\label{sec:training}
We train the classifier offline before using it for online early intrusion detection.
The goal is to automatically learn the spatial-temporal features of raw network flows and utilize those learned features to reliably identify the attack flows as early as possible, that is, after having observed only a small part (e.g., first 2 or 3 packets) of the given flow.

We require a labeled flow dataset for supervised training that contains normal and attack flows. In addition to the labeled flows, the dataset should also have network packets corresponding to the flows.
The majority of the publicly available datasets used for training and evaluating the IDSs have the class imbalance problem~\cite{Ahmad2021}, i.e., the number of examples among the different classes is not similar in the dataset. A classifier trained on an imbalanced dataset typically exhibits poor performance in terms of the overall prediction accuracy.
Therefore, in this work, in order to rectify the effect of class imbalance, we train our classifier with sample weighting, which acts as a coefficient for the loss value computed for each sample (i.e., flow) during the training process. The weight of each sample is based on its class. It is calculated inversely proportional to the class frequencies in the training data. The objective is that the classifier should pay more attention to those samples that belong to an under-represented class.

We prepare our training dataset by processing every packet in the flows using the procedure described in Section~\ref{sec:packet-pre-processing}. 
We denote a flow dataset as $D=\{(F_T^{(j)},y_j)\}$ for $1 \leq j \leq N$, where \textit{N} represents the total number of flows $F_T$ and their corresponding labels $y$.
Since our objective is to train the classifier capable of reliably detecting the attack flow after observing the first few packets out of a given flow, we extend the dataset by cumulatively creating short segments of a flow at different lengths.
The process of extending the training dataset by generating more data (e.g., network flows) from existing data is called \textit{data augmentation}~\cite{Shorten2019}.
We start by creating the shortest segment of a given flow containing only the first packet of the flow; subsequently, we create more segments based on the flow by cumulatively adding more packets with respect to a pre-defined segmentation rate $s_{r}$ such that $0 < s_r < 1$. Segmentation rate $s_{r}$ is a hyper-parameter that is used to calculate the segment size $s_{z} = \lceil s_{r} * T \rceil$ for a given flow, where \textit{T} is the length (i.e. total number of packets) of a flow. 
This parameter value controls the number of segments generated per flow, for instance, more segments per flow will be generated as the value of $s_{r}$ gets smaller.
Suppose we have a flow $F_T^{(j)}=\{P_{1}, P_2,...,P_T\}$, then the set of segments of this flow is as follows: 
\begin{align}
	\{F_{t = k * s_{z}}^{(j)} | k = 1, 2, ..., \lfloor \dfrac{T-1}{s_{z}} \rfloor\}
	\nonumber
\end{align}
where all the segments have the same label $y_j$ as the original flow does.
For example, consider three flows with different lengths: $F_6^{(1)}$, $F_{15}^{(2)}$, and $F_{70}^{(3)}$. We set the segmentation rate $s_{r}$ to 0.25. The segment sizes $s_{z}$ for $F_6^{(1)}$, $F_{15}^{(2)}$, and $F_{70}^{(3)}$ are 2, 4, and 18, respectively. Table~\ref{tbl:flow_segments} lists the segments of the flows generated by the data augmentation process.

\begin{table}[h]
	\caption{Flow segments with respect to the segmentation rate $s_{r}=0.25$}
	\centering
	\label{tbl:flow_segments}
	\begin{tabular}{|l|l|l|}
		\hline
		\textbf{No.} & \textbf{Flows} & \textbf{Flow Segments} \\ \hline
		1 & \multirow{2}{*}{$F_6^{(1)}=\{P_1, P_2,...,P_6\}$} & $\{P_1, P_2\}$ \\ \cline{1-1} \cline{3-3} 
		2 &  & $\{P_1, P_2, P_3, P_4\}$ \\ \hline
		3 & \multirow{3}{*}{$F_{15}^{(2)}=\{P_1, P_2,...,P_{15}\}$} & $\{P_1, P_2, P_3, P_4\}$ \\ \cline{1-1} \cline{3-3} 
		4 &  & $\{P_1, P_2, ..., P_8\}$ \\ \cline{1-1} \cline{3-3} 
		5 &  & $\{P_1, P_2, ..., P_{12}\}$ \\ \hline
		6 & \multirow{3}{*}{$F_{70}^{(3)}=\{P_1, P_2,...,P_{70}\}$} & $\{P_1, P_2, ..., P_{18}\}$ \\ \cline{1-1} \cline{3-3} 
		7 &  & $\{P_1, P_2, ..., P_{36}\}$ \\ \cline{1-1} \cline{3-3} 
		8 &  & $\{P_1, P_2, ..., P_{54}\}$ \\ \hline
	\end{tabular}
\end{table}

Data augmentation is only applied to the flows in the training dataset. The dataset is extended by including the generated flow segments.
We train our early classifier to learn a mapping function $\mathcal{H}:F_t^{(j)} \rightarrow y_j$, where $t \leq T$. In other words, the classifier should be able to predict the class label $y_j$ of a given flow $F_t^{(j)}$ with only first \textit{t} packets.
We have used the categorical cross-entropy loss function and Adam~\cite{Adam2014} optimizer for training our classifier.

\subsection{Monitoring}

\begin{figure*}[htbp]
	\centering 
	\includegraphics[width=\textwidth]{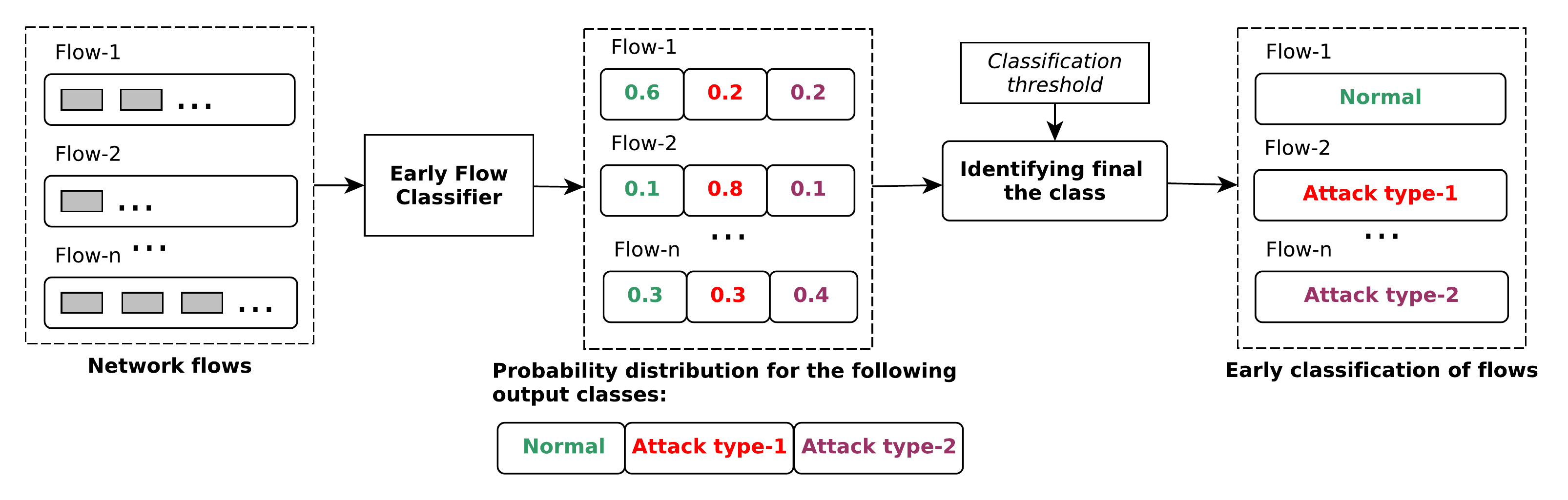}
	\caption{Early classification of flows}
	\label{fig:monitoring}
\end{figure*}

Monitoring high-speed networks in real-time is a challenging task due to the high packet rate. This is the main reason, in our approach, we capture and process only those network packets which are related to the type of attacks we would like to detect. 

The \textit{packet sniffer} module is responsible for network traffic monitoring in real-time. It captures and forwards the inbound and outbound network packets to the flow processing pipeline, as shown in Figure~\ref{fig:approach}. This module is implemented using the \textit{libpcap}\footnote{\url{https://github.com/the-tcpdump-group/libpcap}} library that provides a programming interface to capture packets passing through the network interfaces. The library also supports filters that can be configured to capture only certain packets, for example, packets whose destination port is 80.
Those filters, usually supported by the Operating System kernel, improve performance by reducing the packet filtering process overhead.

We maintain a list of active flows and the predictions corresponding to those flows made by our early flow classifier. Whenever a network flow is updated with a new packet, we use the early flow classifier to get a prediction (i.e., a probability distribution for output classes), as shown in Figure~\ref{fig:monitoring}. 
The final class of a flow is a class that has a higher probability than other classes and the \textit{classification threshold} $\in [0,1)$. 
If none of the class probabilities is higher than a given threshold, our approach will return \textit{Unknown} as the final class.
As we increase the classification threshold, the number of false positives (i.e., a result that indicates a given flow is an attack when it is not) decreases, which improves the classification accuracy but degrades the earliness of the approach.
The threshold is provided by a person such as a network administrator who observes the network traffic and is responsible for taking countermeasures against the attacks  based on the classification results.

\section{Evaluation}\label{sec:evaluation}

In this section, we evaluate the performance of our approach by answering the following research questions:
\begin{itemize}
	\item RQ1: How does our approach perform in classifying complete flows (i.e., flows with all the packets)?

	\item RQ2: How effective is our approach in identifying the class of a given flow in real-time by inspecting only first few packets of the flow?

\end{itemize}
RQ1 investigates the classification performance of our approach; whereas RQ2 evaluates the performance of our approach when deployed in a real-time environment.

We use Scikit-learn~\cite{scikit-learn} library for data pre-processing and Keras~\cite{chollet2015keras} library to build the early flow classifier.
This section describes the dataset and the model architecture used for the evaluation. Lastly, we discuss the results with respect to each research question.

\subsection{Dataset}
We use CICIDS2017~\cite{Sharafaldin2018a} dataset to evaluate the effectiveness of our approach. The dataset is composed of normal and seven types of attack flows (e.g., Hearbleed, Botnet, Web) along with the network packets corresponding to the flows. 
We use a specific part of the dataset that was captured on Thursday, July 6, 2017 and contains network flows related to the following web attacks: (1) \textit{SQL Injection:} an attacker provides a string of SQL commands to be injected in the database; (2) \textit{Cross-Site Scripting (XSS):} an attacker injects a script into the web application code; (3) \textit{Brute Force:} an attacker tries a list of passwords to find the administrator’s password. Table~\ref{tbl:dataset} lists the number of flows and the average flow length (i.e., number of packets) per class in the Thursday dataset.

\begin{table}[h]
	\caption{Labeled Flow Dataset}
	\centering
	\label{tbl:dataset}
	\begin{tabular}{|l|r|r|}
		\hline
		\textbf{Class} & \textbf{Number of Flows} & \textbf{Average Flow length} \\ \hline
		Normal & 27 129 & 124.39 \\ \hline
		Brute Force & 1 507 & 18.43 \\ \hline
		XSS & 652 & 11.48 \\ \hline
		SQL Injection & 21 & 5.71 \\ \hline
	\end{tabular}
\end{table}

We have observed that the header and payload length of 99\% of packets in the dataset are less or equal to 40 and 356 bytes, respectively. To handle the packets with different header and payload lengths, we crop or pad them with zeros at the end to 50 and 400 bytes, respectively, as per transformation step in Section~\ref{sec:packet-pre-processing}. We scale all the packet bytes between 0 and 1 by dividing them by 255. In practice, scaling the input data helps machine learning algorithms converge faster \cite{Sola1997}.

\subsection{Evaluation Metrics}
Evaluating the classification performance of a machine learning-based approach on an imbalanced dataset is a challenging task \cite{Zhu2020}. 
The majority of the existing intrusion detection approaches using machine learning have reported the performance of their approaches using the traditional metrics such as accuracy and F1-score \cite{Ahmad2021}.
These metrics are designed to evaluate the performance of a classifier on balanced datasets. They do not work well when there is a large imbalance in the distribution of the classes in the dataset \cite{Zhu2020}. 
Thus, we evaluate the classification performance of our approach by analyzing the following metrics for each class found in the dataset: precision, recall, false positive rate, \textit{Balanced Accuracy} (BA), and \textit{Bookmaker Informedness} (BM). 
For simplicity, we introduce the metrics here in the context of a binary classification problem where we have only two classes: \textit{positive} and \textit{negative}; however, they can be applied to a multi-classification problem.
\textit{Precision} calculates the percentage of instances identified as positive that were correctly classified, while \textit{recall} (i.e., also known as \textit{detection rate}) computes the percentage of actual positive instances that were correctly classified. \textit{False positive rate} (FPR) (i.e., also known as \textit{false alarm rate}) estimates the proportion of negative observations wrongly predicted as positive over the total number of negative observations.
The BA metric is the arithmetic mean of recall obtained on each class, whereas BM is defined as the probability that the classifier will make a correct decision as opposed to random guessing.
All the metrics mentioned above can have values between 0 and 1. Higher values of precision, recall, BA, and BM and lower values of FPR indicate better classification performance of a classifier. 

To evaluate the earliness of our approach, we define the \textit{earliness} as:
\begin{align}
	Earliness = \dfrac{T-t}{T-1}
\end{align}
where \textit{t} is the minimum number of packets required to correctly predict the class of a given flow, and \textit{T} is the total number of packets in the flow.
Since this metric aims to evaluate the earliness of the prediction instead of its quality, this metric is only applied to those flows that are correctly classified and  $t \leq T$.
The \textit{earliness} value lies between 0 and 1, with extreme values 0 and 1 reached in case a classifier can accurately classify a given flow by analyzing only the first packet and all the packets of the flow, respectively.

\subsection{Model Architecture}
Our model is made of a convolution block. The block contains the following layers in the specified order: 1-D CNN layer using 32 kernels with size 1, valid padding, \textit{ReLU} activation, and bias, layer normalization, and average pooling layer of size 2 with the same padding. We perform global average pooling to flatten the series of feature maps to a fixed-length vector, which is then provided as an input to a fully connected layer of size 64 to get the feature vector. Finally, we use a softmax layer of size 4 to obtain a probability distribution for each class. Based on the probability distribution and the classification threshold, we make the final predictions.   
We set the batch size to 32, which is the number of flows included in a minibatch during neural network training.
The total number of trainable parameters of the model is 16 804. It is trained with 50 epochs on the training dataset.

\begin{figure*}[htbp]
	\centering 
	\includegraphics[width=\textwidth]{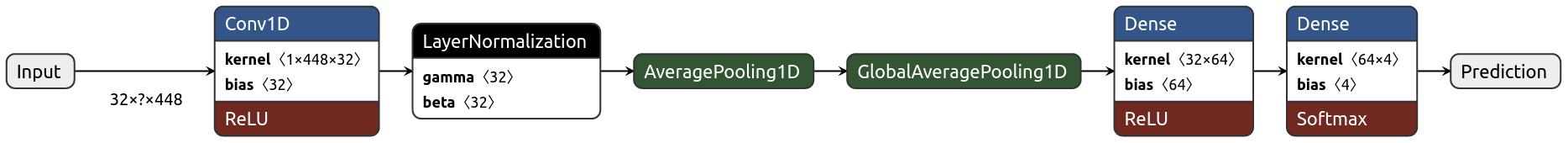}
	\caption{CNN architecture of our early flow classifier}
	\label{fig:model}
\end{figure*}

Figure~\ref{fig:model} portrays the architecture of the model. The label on the arrow from the \textit{Input} to \textit{Conv1D} layer in the figure specifies the dimensions of the input provided to the model. The input has three dimensions: number of flows in a batch, number of packets in a flow, number of bytes or features representing a packet in a flow.
Since we do not fix the number of packets required in a flow to make a prediction, the second dimension in the \textit{32x?x448} label is left open.

\subsection{RQ1: Classification Performance}
The objective of this research question is to investigate classification performance of our approach. 
To answer this question, our classifier is trained and evaluated against the independent test set that is extracted from the dataset. 
We split the dataset into two subsets using the ratio 0.7:0.3: training and test set (see Table~\ref{tbl:train_test_split7030}). 
As one can notice, there is a large imbalance in the distribution of the flow classes; for example, roughly 1291.86 times more flows belong to the normal class than the SQL Injection class.
We have used stratified sampling~\cite{Cochran1977} to split the original dataset. Unlike random sampling, stratified sampling creates the splits by maintaining the same percentage for each class as in the complete dataset.

We used 10-fold cross-validation on the training dataset to fine-tune the hyper-parameter values and model selection. For statistical reasons, the evaluation procedure is repeated 30 times, and every time, we randomly shuffle the dataset to remove any ordering bias before splitting it into training and test set.

\begin{table}[h]
	\caption{Training and test datasets}
	\centering
	\label{tbl:train_test_split7030}
\begin{tabular}{|l|r|r|}
	\hline
	\textbf{Class} & \textbf{Training} & \textbf{Test} \\ \hline
	Normal         & 18 990            & 8 139         \\ \hline
	Brute Force    & 1 055             & 452           \\ \hline
	XSS            & 456               & 196           \\ \hline
	SQL Injection  & 15                & 6             \\ \hline
\end{tabular}
\end{table}

We have augmented the training dataset using the segmentation rate $s_{r}=0.1$. Table~\ref{tbl:aug_flows7030} lists the number of flows we obtained by applying the data augmentation technique described in Section~\ref{sec:training}. 

\begin{table}[h]
	\caption{Number of flows after augmenting the training dataset with $s_{r}=0.1$}
	\centering
	\label{tbl:aug_flows7030}
	\begin{tabular}{|l|r|r|}
		\hline
		\textbf{Class} & \textbf{Original} & \textbf{Augmented} \\ \hline
		Normal         & 18 990            & 92 468            \\ \hline
		Brute Force    & 1 055             & 4 625              \\ \hline
		XSS            & 456               & 1 866              \\ \hline
		SQL Injection  & 15                & 82                 \\ \hline
	\end{tabular}
\end{table}

A perfect IDS has a 1.0 recall at 0.0 FPR for every class, which means that it can identify all flows correctly without any miss-detection. Nevertheless, in reality, such flawless IDSs are empirically not feasible or very difficult to attain in a real-time environment because of the complexity and large volume of network traffic.
Table~\ref{tbl:evaluation} shows the achieved performance of our early flow classifier on the test set.
Our approach gives the highest detection rate or recall of 0.911 at a FPR of 0.008 for the \textit{XSS} attack type among all the other attack types. In other words, our approach correctly identifies 91.1\% of the \textit{XSS} attack flows in the test dataset and wrongly identifies less than 1\% of other types of flows as \textit{XSS} attack flows.
Our approach has performed well also for the other types of attack, even though the number of training flows for the attack types is low. For example, the number of training flows for \textit{Brute Force} and \textit{XSS} attack types are only 5.1\% and 2.2\% of the total number of original training flows. One can notice that the approach has performed poorly for the \textit{SQL Injection} attack type. The main reason is that the number of samples of the attack type is significantly small (i.e., 0.07\% of the total number of training samples). Thus, the model has a limited capacity to learn the attack type.
Overall, our approach performed well and attained 0.803 balanced accuracy.

\begin{table}[h]
	\caption{Classification performance}
	\centering
	\label{tbl:evaluation}
\begin{tabular}{|l|l|l|l|l|}
	\hline
	\textbf{Class} & \textbf{Precision} & \textbf{Recall} & \textbf{FPR}   & \textbf{BM} \\ \hline
	Normal         & 0.996              & 0.944           & 0.054 & 0.891       \\ \hline
	Brute Force    & 0.720              & 0.828           & 0.051 & 0.778       \\ \hline
	XSS            & 0.754              & 0.911           & 0.008 & 0.904       \\ \hline
	SQL Injection  & 0.343              & 0.528           & 0.003 & 0.525       \\ \hline
\end{tabular}
\end{table}

\subsection{RQ2: Earliness Performance}
This research question aims to study the performance of our approach in detecting attacks as early as possible in a real-time environment. In our opinion, a real-time IDS should satisfy the following two requirements. First, the IDS should be able to process the data (i.e., network packets) as fast as it is being produced under normal circumstances. Second, the minimum number of packets required to accurately predict the class of a given flow (MNP) should be less than the total number of packets in the flow.

To answer this question, we ran a replay session where we reproduced the network traffic captured in the dataset against our approach to emulate a real-time environment. 
Our approach and the software that replayed the traffic ran on different machines. Each machine featured an Intel Core i9-10900X CPU, 64 GB of memory, RTX 3090 graphics card, and Ubuntu 20.04 Operating System. The machines were connected via a 1Gb Ethernet connection in an isolated environment to reduce network latency.

The replay session ran for 29 146 seconds and re-transmitted \mbox{9 322 025} packets.  We configure the packet filtering module to forward only those packets whose destination or source port is 80 since we are interested in detecting web attacks.
During the replay session, we monitor packet inter-arrival times, processing times required by our approach to make predictions, and the minimum number of packets required (MNP) to predict a flow class accurately.

The average packet inter-arrival time of all the packets and the filtered packets were 3.13 and 15.96 milliseconds, respectively. 
On average, our approach was able to make a prediction in 0.04 milliseconds per packet, for example, if a flow has 4 packets, our approach would take 0.16 milliseconds to predict its class.
This shows that approach is able to process the network packets faster than they are being produced and satisfies the first requirement.
Table \ref{tbl:earliness} shows the earliness, MNP, and the average flow length per class. The results show that our approach can detect the class of a given flow by inspecting roughly only 1 to 3 packets. 
In summary, our approach is feasible to be deployed in a real-time environment to detect attacks by inspecting a few packets.

\begin{table}[h]
	\caption{Earliness metric and the average minimum number of packets required (MNP) to predict the flow class}
	\centering
	\label{tbl:earliness}
	\begin{tabular}{|l|r|l|r|}
		\hline
		\textbf{Class} & \textbf{Earliness} & \textbf{MNP}  & \textbf{Average Flow length} \\ \hline
		Normal         & 0.991              & 2.11 & 124.39                       \\ \hline
		Brute Force    & 0.936              & 1.17 & 18.43                        \\ \hline
		XSS            & 0.917              & 1.00 & 11.48                        \\ \hline
		SQL Injection  & 0.509              & 2.80 & 5.71                         \\ \hline
	\end{tabular}
\end{table}

\section{Related Work}\label{sec:related-work}

Recently, a number of deep learning-based IDS approaches have been proposed. 
Most of these approaches (e.g., \cite{Gu2021,Umer2017,Marir2018,Vinayakumar2019,Malaiya2019,Andresini2020, Rajagopal2021}) rely on flow-based statistical features extracted by analyzing all the packets in a given flow such as total bytes, packets count, IP addresses and ports numbers. 
In contrast, the proposed approach aims to extract relevant features from raw network traffic data that can be used to reliably detect attack flows by analyzing the partial information already available of the flows during the early phase of attacks.
In this section, we focus on some of the most important and recent related works on IDS that use machine learning to classify network attacks by extracting the relevant features from raw network traffic data.

Zhang et al. \cite{Zhang2019a} proposed an intrusion detection network based on a convolutional neural network, named parallel cross convolutional neural network (PCCN). 
They use the network traffic data to extract features, but they restrict the number of packets in a flow to 5.
The authors mention that the PCCN network structure meets the real-time requirements of network intrusion detection; however, they neither further discuss nor evaluate this aspect of the approach.

Zhu et al.~\cite{Zhu2021} presented a hierarchical network intrusion detection model based on unsupervised clustering using deep auto-encoder and gaussian mixture model. 
The proposed model comprises two sub-models: the first sub-model detects abnormal traffic in real-time, and the second identifies the attack categories of abnormal traffic detected by the first one.
They employ the feature processing method from PCCN approach~\cite{Zhang2019a} to obtain features for their intrusion detection model.
The authors state that essential features are extracted based on the first few packets, which guarantee real-time network intrusion detection. However, they neither discuss how the approach achieves real-time intrusion detection nor evaluate this aspect of the approach.
Further, they report the performance of their approach in terms of accuracy, F1-scores, and AUC averaged over all the classes, which could be misleading when the class distribution is imbalanced~\cite{Zhu2020}.

Zhang et al. \cite{Zhang2019} proposed a network intrusion detection model that integrates CNN and LSTM neural network structures to learn the spatial and temporal features of flows. 
Similar to our approach, they use the network traffic data to extract features. However, they restrict the number of packets in a flow to 10, whereas we do not limit the number of packets in a flow and, in addition, analyze all the packets available in order to make an informed prediction.
Further, they also report the performance of their approach with respect to the accuracy, F1-scores, and AUC averaged over all the classes, which could be misleading when the class distribution is imbalanced~\cite{Zhu2020}.

Zhang et al. \cite{Zhang2019b} proposed a multiple-layer representation learning model for network intrusion detection by combining convolutional neural networks (CNN) with gcForest. They propose a new data encoding scheme based on P-Zigzag to encode a network packet into a two-dimensional gray-scale image for classification. 
In contrast to our approach, this approach classifies packets instead of flows.

L\'{o}pez-Vizca\'{\i}no et al. \cite{LopezVizcaino2019} defined the early intrusion detection problem by grouping network packets into data flows, where each flow is labeled as an attack or normal traffic depending on the intent of its packets. The ideas and concepts in this work are very relevant to our work.
The authors propose a new time-aware metric, named ERDE, where accurate predictions are penalized if they are made after a certain measuring point \textit{o} that is defined manually. This metric was initially proposed to measure the early detection of depressed individuals based on their posts on a social network. In contrast, our non-parametric \textit{earliness} metric is designed specifically for network flows. 
The metric value ranges from 0 to 1, with extreme values 0 and 1 reached if a classifier can accurately classify a given flow by analyzing only the first packet and all the flow packets, respectively.
In comparison to ERDE, we consider our metric to be more informative, comparable, and intuitive.
For evaluation, the authors above divide every flow into ten chunks containing 10\% of the packets for each flow. A set of classifiers (i.e., Random Forest, J48, JRip and PART) analyzes each chunk of flows sequentially and, it can produce three outputs: attack, normal or delay. The objective is to detect an attack using as few chunks as possible.
They utilize the feature extraction method from~\cite{Mirsky2018} that extracts traffic statistics, such as source port, IP, and MAC addresses, from every new packet transmitted over a network channel. Although the authors define the early intrusion detection problem in terms of network flows, they do not explain how do they utilize the features extracted using a method (that does not consider network flows and processes each packet independently) in order to predict the class of a given flow.
In contrast, we describe packet pre-processing steps and the features used for early classification in Section~\ref{sec:flow-processing-pipeline} and \ref{sec:training}, respectively in detail.
The authors conclude that machine learning models do not perform well when they are used for early intrusion detection; however, our results show that our approach can identify attacks with a high degree of accuracy by analyzing the first few packets of a given flow.

In summary, to the best of our knowledge, the existing intrusion detection approaches can detect a certain attack by inspecting the complete information related to the attack. This means that a system would only be able to detect an attack after it has been executed on the system under target and might have caused damage to the system.
In contrast, our end-to-end early intrusion detection approach can reliably detect attacks by analyzing the partial information already available in the early phase of attacks.

\section{Threats to Validity}\label{sec:threats-to-validity}
The main threat to external validity is that we used only one dataset in our evaluation. Therefore, the results of our experiment may differ for other datasets that have different types of attack classes. 
To the best of our knowledge, most of the publicly available datasets, excluding CSE-CIC-IDS2018\footnote{\url{https://www.unb.ca/cic/datasets/ids-2018.html}} that is just a successor of CICIDS2017, either are outdated and/or lacking raw network traffic data \cite{Ahmad2021,Sharafaldin2018a}.
Our approach can extract relevant features from raw network traffic data instead of relying on a manual feature selection process; thus, it can easily be applied to other datasets.
It will be the future work for us to perform additional experiments using different datasets such as CSE-CIC-IDS2018 to mitigate this threat.

Another threat to external validity is that the evaluation might seem subjective because we have not compared our approach with other IDS approaches from the literature. However, as discussed in Section~\ref{sec:related-work}, we could not find any approach similar to ours that extracts the relevant features from raw network traffic data in an end-to-end manner instead of relying on manual or flow-based statistical features and detects network attacks as early as possible.

An internal threat to our work is that we have considered only three attack classes. For example, we have not experimented with other types of attacks such as Hearbleed and Botnet. 
As we have mentioned in Section~\ref{sec:early-flow-classification}, we train a simple DNN model (i.e. a model that has a relatively small number of trainable parameters) to detect only certain types of attacks in order to obtain good accuracy and a reasonable prediction time.

\section{Conclusion}\label{sec:conclusion}
In this paper, we have presented an end-to-end early intrusion detection system to prevent network attacks in real-time before they could cause any more damage to the system under attack. 
We have utilized a CNN-based classifier for early attack identification. The network is trained in a supervised manner to extract relevant features from raw network traffic data instead of relying on a manual feature selection process used in most related approaches.
Additionally, we have introduced a new metric, earliness, to evaluate the earliness of the predictions made by our approach.
We have empirically evaluated our approach on the CICIDS2017 dataset. The results show that our approach identifies attacks with a high degree of accuracy by analyzing roughly only 1 to 3 packets. Our approach has achieved overall 0.803 balanced accuracy.
For our future work, we aim to evaluate our approach on different datasets. Further, we plan to explore different neural network architectures to study their relative performance on early attack identification.
We are also planning to employ our approach to protect nodes on a IoT (Internet of Things) network using \textit{Tensorflow Lite for Microcontrollers} \cite{David2020} library that allows to run pre-trained machine learning models on microcontrollers units with only KiloBytes (KB) of memory.

\section*{Acknowledgment}
This work was made possible with funding from the European Union’s Horizon 2020 research and innovation programme, under grant agreement No. 957212 (VeriDevOps). The opinions expressed and arguments employed herein do not necessarily reflect the official views of the funding body.
\IEEEtriggeratref{35}
\bibliographystyle{IEEEtran}

\bibliography{paper}

\end{document}